\documentclass[12pt]{article}
\usepackage[cp1251]{inputenc}
\usepackage[english]{babel}
\usepackage{amsfonts}

\begin{document}
\begin{center}
{\Large
Commutative Rings of Differential Operators
Corresponding to Multidimensional Algebraic Varieties
}
\end{center}
\begin{center}
{A.E. Mironov}
\end{center}

\section{ Introduction}

In this article we construct commutative rings of multidimensional
($N\times N$)-matrix differential operators whose
common eigenfunctions and eigenvalues are parametrized
by points of a~spectral variety~
$Y^k$,
the intersection of some smooth hypersurfaces
$Y_{a_1}\cap\dots\cap Y_{a_k}$
in a~principally polarized abelian variety
~$X^g$
of dimension
$g$, $k<g-1$.
The hypersurface
$Y_{a_j}$
is the translate
of a~theta-divisor $Y\subset X^g$
by an~element $a_j\in X^g$.
The number
$N$
equals
$rd_g$,
where
$d_g$
is the $g$-fold self-intersection index  of the hypersurface~$Y$.
Denote by
$Q^j$
the variety
$Y^j\cap Y$.
Below we suppose that the variety~$Y^j$
intersects
$Y_{a_{j+s}}$
and~$Y$, $j+s\leq k$, transversally.
We assume that
$Y^j$
and
$Q^j$
are smooth and irreducible and the collection
$a_1,\dots, a_k$
is in  general position (i.e., belongs to some
open everywhere dense set in
$X^g\times\dots\times X^g$).

These commutative rings relate to some analog of the
Kadomtsev--Petviashvili hierarchy to be indicated in this article.

Our main result is the following

{\bf Theorem~1.} {\sl
There is an~embedding
$L_k$
of the ring of meromorphic functions on the variety~
$Y^k$
with a~pole on
$Q^k$
into the ring of $(N\times N)$-matrix differential operators in
$g-k$
variables whose coefficients are analytic in a~neighborhood of~$0$:
$$
L_k:{\cal A}_k\rightarrow {\rm Mat}(N,g-k).
$$
The range of the embedding is a~commutative ring
of $(g-k)$-dimensional matrix differential operators.
}

Using the Riemann--Roch--Hirzebruch theorem, we can demonstrate that
the number~
$d_g$,
and in consequence ~$N$,
is a~multiple of~$g!$.

The operators
$L_k({\cal A}_k)$
have rank~$r$.
This means that to each point of ~
$Y^k$
there corresponds~
$r$
linearly independent eigenfunctions.

The two-dimensional operators
$L_k({\cal A}_k)$
with doubly periodic coefficients are finite-gap at every energy
level
$E$;
i.e., the Bl\^och vector-functions
(eigenfunctions of both the operators
$L_k(\lambda)$, $\lambda\in {\cal A}_k$,
and the translation operators by periods)
are parametrized by a~Riemann surface of finite genus
defined in the spectral surface by the equation
$\lambda=E$.

If the dimension of~
$Y^k$
equals~2 then, using the adjunction formula and
the Lefschetz embedding theorem, we can demonstrate that
the Kodaira dimension of the spectral surface
$Y^k$
equals~2; i.e., this surface is a~surface of general type.

We prove Theorem~1, using Nakayashiki's results~[1]
(see also~[2])
who constructed an~embedding of the ring of meromorphic functions on
$X^g$
with a~pole on
$Y$ into the ring of $g$-dimensional
($N\times N$)-matrix differential operators. The
($2\times 2$)-matrix operators of this kind in two variables
(the Nakayashiki operators) were studied in the author's
articles~[3,\,4].
In particular, it was proven in~[4] that there are no two-dimensional
real Nakayashiki operators that are finite-gap at every energy level
and have doubly periodic coefficients, but
there exist two-dimensional real Nakayashiki operators
with singular doubly periodic coefficients which are finite-gap at
every
energy level. In~[4] we also indicated smooth real Nakayashiki
operators,
including a~second-order operator~$H$
whose diagonal is constituted by Schr\"odinger operators
in doubly periodic magnetic fields with doubly periodic
potentials of the form
$$
(\partial_{y_1}-A_1)^2+
(\partial_{y_2}-A_2)^2+u(y),\quad  y=(y_1,y_2).
$$
The magnetic Bl\^och vector-functions of the operator
$H$
(the common eigenfunctions of~
$H$
and the magnetic translation operators~
$T^*_j$, $T^*_j\varphi(y)=\varphi(y+e_j)\exp(2\pi y_j)$,
$j=1,2$, where $e_j$
are the periods) are parametrized
by a~Riemann surface of finite genus at each energy level. This
property is an~analog of the finite-gap property
at each energy level for operators with doubly periodic coefficients.

In the particular case when
$g=3$, $r=1$,
and the spectral surface is a~theta-divisor
Theorem~1 was proven by Nakayashiki~[1].

Rothstein~[5] constructed another example of commuting
matrix differential operators. In this example $g=5$, $r=1$,
the size $N$ of the matrices equals~$5$,
and the spectral surface is the Fano surface.

Let us recall  Krichever's construction~[6] of commuting
ordinary differential operators of rank~1.
Suppose that
$\Gamma$
is a~Riemann surface of genus~
$g$,
$P=p_1+\dots +p_g$
is a~nonspecial positive divisor on
$\Gamma$,  $\infty$
is a~point on~
$\Gamma$
other than the points of ~
$P$,
$k^{-1}$
is a~local parameter at~$\infty$,
and
$k^{-1}(\infty)=0$.
There is a~Baker--Akhiezer function
$\psi(p,x)$, $p\in\Gamma$,
meromorphic on
$\Gamma\backslash \infty$ and
whose set of poles coincides with
$P$
and is independent of~$x$; moreover, the function~
$\psi\exp(-kx)$
is analytic in a~neighborhood of~
$\infty$.
For every meromorphic function
$f(p)$
on~
$\Gamma$
with a~sole pole at~
$\infty$
there is a~unique differential operator~
$L(f)$
such that
$$
L(f)\psi=f\psi.
$$
The operators
$L(f)$
commute pairwise for different~
$f$.
Hence, we obtain a~relation between the spectral data
of the commuting Burchnall--Chaundy--Krichever operators and the
spectral
data of the operators~
$L_k({\cal A}_k)$:
$$
\{\Gamma,\infty,P,f\}
\longleftrightarrow
\bigl\{Y^k,Q^k,Q^k_c,\lambda\bigr\},
$$
where
$Q_c^k=Y^k\cap Y_c$ and  $c\in X^g$
is some nonzero element.

As in the one-dimensional case we can construct operators
$L_{\alpha}$
whose coefficients depend on time and satisfy some evolution
equations.

{\bf Theorem~2.} {\sl
There is a~multidimensional analog of the Kadomtsev--Petviashvili
hierarchy
$$
[\partial_{t_{\alpha}}-L_{\alpha},
\partial_{t_{\beta}}-L_{\beta}]=0,
$$
where
$L_{\alpha}$
and
$L_{\beta}$
are $(N\times N)$-matrix differential operators in~
$g-k$
variables whose coefficients depend on~
$t_{\alpha}$
 and
$t_{\beta}$
with
$\alpha$
and
$\beta$
varying in some countable set of indices.
}

As was already mentioned in~[4], the coefficients of the Nakayashiki
operators
cannot satisfy evolution equations of the
Kadomtsev--Petviashvili hierarchy type.

In Section~2 we introduce vector theta-functions which determine
sections of holomorphic vector bundles of rank~
$r$
over an~abelian variety~
$X^g$.
For
$r=1$
the vector theta-functions coincide with the classical Riemann
theta-functions.
Using vector theta-functions, we can write down explicitly
sections of holomorphic vector bundles over Riemann surfaces. If
$X^g$
is the Jacobi variety of a~Riemann surface~
$\Gamma\subset X^g$
then the restriction of a~vector theta-function to~
$\Gamma$
is a~section of the vector bundle over~
$\Gamma$
of rank
$r$
and degree
$rsg$,
where
$s$
is some natural number and $g$ is the genus of~$\Gamma$.

In Section~3, using the Fourier--Mukai transform~[7],
we introduce the Baker--Akhiezer module over the ring of
differential operators whose elements are expressed in terms of
vector theta-functions. Theorems~1 and 2 ensue from Theorem~3
which claims that the Baker--Akhiezer module is free.

The author is grateful to I.~A. Ta\u\i manov for useful
discussions and remarks.

\section{ Vector Theta-Functions}

In this section we indicate the coefficients of the Fourier series
expansion
of vector theta-functions. In~Lemma~1 we find the dimension
of the space of vector theta-functions.

Denote by
$X^g={\Bbb C}^g/\{{\Bbb Z}^g+\Omega {\Bbb Z}^g\}$
a~principally polarized complex abelian variety, where
$\Omega$
is a~symmetric ($g\times g$)-matrix with
${\rm Im}\Omega>0$.
Given nondegenerate pairwise commuting ($r\times r$)-matrices
$A_j$, $j=1,\dots,g$,
introduce the set of matrix functions (multipliers) on
${\Bbb C}^g$:
$$
e_{n+\Omega m}(z)=
\exp(-s\pi i\langle m,\Omega m\rangle -
2s\pi i\langle m,z\rangle)A_1^{m_1}\dots A_g^{m_g},
$$
where
$m,n\in{\Bbb Z}^g$, $\langle m,z\rangle=m_1z_1+\dots+m_gz_g$,
and
$s$ is some natural number.
It is easy to check that these functions satisfy the equalities
$$
e_{\lambda}(z+\lambda')e_{\lambda'}(z)=
e_{\lambda'}(z+\lambda)e_{\lambda}(z)=
e_{\lambda+\lambda'}(z),\quad
\lambda,\lambda'\in {\Bbb Z}^g+\Omega{\Bbb Z}^g.
$$
An~arbitrary collection of matrix functions satisfying these
equalities determines a~vector bundle of rank~
$r$
over~$X^g$
which is obtained by factoring
${\Bbb C}^g\times{\Bbb C}^r$
by the action of the lattice
${\Bbb Z}^g+\Omega {\Bbb Z}^g$:
$$
(z,v)\sim
(z+\lambda,e_{\lambda}(z)v), \quad v\in{\Bbb C}^r.
$$
Global sections are given by vector-functions on
${\Bbb C}^g$
with the periodicity properties
$$
f(z+\lambda)=e_{\lambda}f(z).
$$
A~{\it vector theta-function of rank
$r$
and degree}
$s$
is a~vector-function
$$
\theta^{r,s}(z)=\bigl(\theta^s_1(z),\dots,\theta^s_r(z)\bigr)^{\top},
\quad z\in{\Bbb C}^g,
$$
with entire components which possesses the property
$$
\theta^{r,s}(z+\Omega m+n)=
\exp(-s\pi i\langle m,\Omega m\rangle -
2s\pi i\langle m,z\rangle)A_1^{m_1}\dots A_g^{m_g}
\theta^{r,s}(z).
$$
By periodicity,
$\theta^{r,s}$
expands in the series
$$
\theta^{r,s}=\sum\limits_{l\in{\Bbb Z}^g}
\exp(2\pi i\langle l,z\rangle)a_l,
\quad a_l=\bigl(a_l^1,\dots, a_l^r\bigr)^{\top}\in{\Bbb C}^r.
$$
Find the recurrent relations for the coefficients
$a_l$:
$$
\theta^{r,s}(z+\Omega e_j)=
\sum\limits_{l\in{\Bbb Z}^g}
\exp(2\pi i\langle l,\Omega e_j\rangle)
\exp(2\pi i\langle l,z\rangle)a_l
$$
$$
=\sum\limits_{l\in{\Bbb Z}^g}
\exp(-s\pi i\Omega_{jj})
\exp(2\pi i\langle l-se_j,z\rangle)A_ja_l;
$$
consequently,
$$
a_{l+se_j}=
\exp(s\pi i\Omega_{jj}+2\pi i\langle l,\Omega e_j\rangle)
A_j^{-1}a_l,
$$
where
$
e_j=(0,\dots,1,\dots,0)^{\top}
$
(with $1$ at the $j$th place).
It follows from the last formula that
$\theta^{r,s}$
is determined by the coefficients
~$a_l$,
where the components of~
$l$
vary within
$0\leq l_{\alpha}\leq s-1$;
therefore, the dimension of the space of vector theta-functions
does no exceed~
$rs^g$.
Show that for every choice of
$a_l$, $0\leq l_{\alpha}\leq s-1$,
the series for the vector theta-function
$\theta^{r,s}$
converges. To this end, rewrite it as follows:
$$
\theta^{r,s}(z)=\sum\limits_{l_0} \sum\limits_{l\in{\Bbb Z}^g}
\exp(2\pi
i\langle l_0+sl,z\rangle)a_{l_0+sl},
$$
where the components of
$l_0$
vary between $0$ and~$s-1$.
The above recurrent relations can be resolved explicitly:
$$
a_{l_0+sl}=
\exp(s\pi i\langle l,\Omega l\rangle+
2\pi i\langle l_0,\Omega l\rangle)
A_1^{-l_1}\dots A_g^{-l_g}
a_{l_0}.
$$
Put
$$
\theta_{a_{l_0}}^{r,s}=
\sum\limits_{l\in{\Bbb Z}^g}
\exp(s\pi i\langle l,\Omega l\rangle+
2\pi i\langle l_0,\Omega l\rangle+
2\pi i\langle l_0+sl,z\rangle)
A_1^{-l_1}\dots A_g^{-l_g}
a_{l_0}.
$$
Then
$$
\theta^{r,s}=\sum\limits_{l_0}\theta_{a_{l_0}}^{r,s}.
$$
Denote by
$C_j$
the greatest of the two numbers
$\bigl\| A_j^{-1}\bigr\|$
and
$\| A_j\|$.
Then the norm of each summand in the series for
$
\theta_{a_{l_0}}^{r,s}
$
does not exceed
$$
|\exp(s\pi i\langle l, \Omega l\rangle+2\pi i
\langle l_0, \Omega l\rangle+2\pi i
\langle l_0+sl, z\rangle)|C_1^{|l_1|}
\dots C_g^{|l_g|}\|a_{l_0}\|;
$$
consequently, by positive definiteness of~
${\rm Im}\,\Omega$
this series converges absolutely.
We obtain the following

{\bf Lemma 1.} {\sl
The dimension of the space of vector theta-functions of degree
$s$
and rank~
$r$
equals
$rs^g$.
}

We give an~example of the matrices
$A_j$.
We take
$A_1$
to be some matrix with nondiagonal Jordan form and the remaining
$A_j$
to be polynomials in~
$A_1$.
If the matrices
$A_j$
have diagonal Jordan forms
then the bundle corresponding to the collection
$A_j$
is the direct sum of line bundles.

\section { Commuting Operators}

In this section we state Nakayashiki's theorem~[1]
in the particular case of holomorphic vector bundles of rank
$r\geq 1$
invariant under translations by elements of~
$X^g$
which  we need below. Using the Fourier--Mukai transform
of these bundles, we introduce the Baker--Akhiezer modules~
$M_c^j$
over the ring
${\cal D}_j$
of differential operators.
We show in Corollary~2 that
the restriction map of functions in~
$M_c^j$
to the variety
$Y^{j+1}\subset Y^j$
determines an~epimorphism~
$ M_c^j\rightarrow M_c^{j+1}$.
In~Theorem~3 we prove that the ${\cal D}_j$-module
$M_c^j$
is free. In~Corollary~3 we show that the coefficients of the
operators
$L_k({\cal A}_k)$
satisfy some evolution equations.

Denote by
${\rm Pic}^0 (X^g)$
the Picard variety of~
$X^g$.
In our case
$X^g$
and
${\rm Pic}^0 (X^g)$
are isomorphic.
Denote by
${\cal P}$
the Poincar\'e bundle over
$X^g\times {\rm Pic}^0(X^g)$.
The sections of~
${\cal P}$
under the lift to~
${\Bbb C}^g\times{\Bbb C}^g$
are determined by functions
$f(z,x)$
such that
$$
f(z+\Omega m_1+n_1,x+\Omega
m_2+n_2)= \exp(-2\pi i(\langle m_1,x\rangle + \langle
m_2,z\rangle))f(z,x),
$$
where
$m_j$, $n_j\in {\Bbb Z}^g$.

Let
$Y$
represent the zeros of some theta-function
$\vartheta$
(of rank~1)
of degree~$s$:
$$
\vartheta(z+\Omega m+n)=
\exp(-s\pi i\langle m,\Omega m\rangle -
2s\pi i\langle m,z\rangle)\vartheta(z).
$$
Denote by
${\cal L}_c$
the holomorphic vector bundle over~
$X^g$
whose sections are given by the vector-functions
$f(z)$
of rank~$r$ on~
${\Bbb C}^g$
with the property
$$
f(z+\Omega m+n)=\exp(-2\pi i
\langle m,c\rangle )A_1^{m_1}\dots A_g^{m_g}f(z),
\quad m,n\in {\Bbb Z}^g, \ c\in{\Bbb C}^g.
                                               \eqno{(1)}
$$
Observe that the bundle
${\cal L}_c$
is invariant under translations by the elements of~
$X^g$.
Let
${\cal L}$
be the space of global sections of the bundle
${\cal L}_0$
with a~pole on~$Y$ and let $\pi$ be the projection
$X^g\times {\rm Pic}^0(X^g)\rightarrow X^g$.
Denote by
$F(Y,{\cal L}_0)(U)$
the space of meromorphic sections of the bundle
$\pi^*{\cal L}_0\otimes {\cal P}$
over
$X^g\times U$
with a~pole on
$Y\times U$,
where
$U$
is an~open subset in
${\rm Pic}^0(X^g)$.
For a~given
$x\in U$
the space
$F(Y,{\cal L}_0)(U)$
coincides with the space
$\bigcup\nolimits_{j=1}^\infty {\rm H}^0(X^g,{\cal L}_{x}(jY))$.
We sometimes denote by
${\cal L}_x(jY)$
the bundle
${\cal L}_x\otimes [jY]$,
where
$[jY]$
is the line bundle associated with the divisor~$jY$.
For simplicity we denote vector bundles and
the corresponding bundles of analytic sections by the same symbol.
We identify the space
${\rm H}^0(X^g,{\cal L}_{x}(jY))$
with the space of global sections of the bundle
${\cal L}_{x}$
with a~pole on~$Y$;
moreover, the order of the pole does not exceed~$j$.

The space
$F(Y,{\cal L}_0)(U)$
is the {\it Fourier--Mukai transform\/} over~$U$
of the space~${\cal L}$.

The covariant differentiation operators act on
$F(Y,{\cal L}_0)(U)$:
$$
\nabla_j=\partial_{x_j}-
\frac{1}{s}\partial_{z_j}\log\vartheta(z)
:F(Y,{\cal L}_0)(U)
\rightarrow
F(Y,{\cal L}_0)(U),
$$
$$
\nabla_k\nabla_j=\nabla_j\nabla_k,\quad
k,j=1,\dots,g,
$$
which furnish
$F(Y,{\cal L}_0)(U)$
with the structure of a~module over the ring
${\cal O}_U[\nabla_1,\dots,\nabla_g]$,
where
${\cal O}_U$
is the ring of analytic functions on~$U$.
It follows from the construction that
$F(Y,{\cal L}_0)(U)$
is also a~module over the ring
${\cal A}_0$
of meromorphic functions on~
$X^g$
with a~pole on~$Y$.

Denote by
${\cal D}_g$
the ring of differential operators
${\cal O}_g[\partial_{x_1},\dots,\partial_{x_g}]$,
where
${\cal O}_g$
is the ring of analytic functions in the variables
$x_1,\dots,x_g$
which are defined in a~neighborhood of~
$0\in{\Bbb C}^g$.
In~[1] Nakayashiki introduced the Baker--Akhiezer module
$M_c=\bigcup\nolimits_{n=1}^{\infty}M_c(n)$
over the ring ${\cal D}_g$ of differential operators,
where
$$
M_c(n)=
\Biggl\{
f(z,x) \exp
\Biggl(-\sum\limits_{j=1}^g \frac{x_j}{s}\,
\partial_{z_j}\log\vartheta(z)\Biggr),\
 f(z,x)\in {\rm H}^0(X,{\cal L}_{c+x}(nY))
\Biggr\}.
$$
We need one more ${\cal D}_g$-module
$$
{\cal D}_gM_c(n)=\Bigl\{\sum d\varphi,\ d\in{\cal D}_g,
\ \varphi\in M_c(n)\Bigr\}.
$$
We can express the elements of
$M_c$
in terms of vector theta-functions.  Every vector-function in~$M_c$
is representable as the sum of vector-functions of the form
$$
g(x)\frac{\theta^{r,sn}(z+\frac{x+c}{sn})}
{\vartheta^n(z)}
\exp\Biggl(-\sum\limits_{j=1}^g\frac{x_j}{s}\,
\partial_{z_j}\log\vartheta(z)\Biggr),
$$
where
$g(x)\in{\cal O}_g$
and
$\theta^{r,sn}$
is some vector theta-function.

The following theorem is proven in~[1]:

{\bf Nakayashiki's Theorem.} {\sl
For $c$ in general position,
$M_c$
is a~free ${\cal D}_g$-module
of rank~$N$. The equality
$M_c={\cal D}_gM_c(g)$
is valid.
}

The equality
$M_c={\cal D}_gM_c(g)$
means that the
${\cal D}_g$-module
$M_c$
is generated by the elements of~
$M_c(g)$.

Fix a~basis
$\Phi_c=(\phi_{1,c}(z,x),\dots,\phi_{N,c}(z,x))^{\top}$
for the ${\cal D}_g$-module~$M_c$.
Sometimes, like in the following corollary, by
$\Phi_c$
we mean the matrix function with
$N$
rows and
$r$
columns, since each component
$\phi_{j,c}(z,x)$
is itself a~vector-function of size~$r$.

{\bf Corollary 1 \rm [1].} {\sl
There is a~ring embedding
$$
L_0:{\cal A}_0\rightarrow {\rm Mat}(N,g)
$$
defined by the equality
$$
L_0(\lambda)\Phi_c=\lambda\Phi_c, \quad \lambda\in{\cal A}_0.
$$
The range of the embedding is a~commutative ring
of $g$-dimensional matrix differential operators.
}

We turn to the construction.
In fact, we prove a~stronger version of Theorem~1.
We assume that the hypersurface
$Y_{a_j}$
can be not only a~translate of~
$Y$
but also a~translate of some smooth hypersurface linearly
equivalent to~$Y$;
i.e.,
$Y_{a_j}$
is the set of zeros of some theta-function of degree
$s$
with a~translation:
$$
Y_{a_j}=\{z\in X^g,\ \vartheta_j(z-a_j)=0\}.
$$
Denote by
${\cal L}_c^k$
the line bundle over~$Y^k$
whose sections are given by the functions
$f(z)$
on~
$Y^k\subset{\Bbb C}^g$
with property~(1).

Introduce the Baker--Akhiezer module
$M_c^k=\bigcup\nolimits_{n=1}^{\infty}M_c^k(n)$
over
${\cal D}_{g-k}$,
where
$$
M_c^k(n)=
\Biggl\{
f(z,x)
\exp
\Biggl(
-\sum\limits_{j=1}^g
\frac{x_j}{s}\,
\partial_{z_j}\log\vartheta(z)
\Biggr),\
f(z,x)\in
{\rm H}^0\bigl(Y^k,{\cal L}_{c+x}^k(nQ^k)\bigr)
\Biggr\}.
$$

{\bf Theorem~3.} {\sl
For
$c$
in general position,
$M_c^k$ is a~free ${\cal D}_{g-k}$-module
of rank~$N$.
}

To prove this theorem, we need

{\bf Lemma 2.} {\sl
The restriction map
$$
\pi_j:
{\rm H}^0\bigl(Y^j,{\cal L}_{c+x}^j(nQ^j)\bigr)
\rightarrow
{\rm H}^0\bigl(Y^{j+1},
{\cal L}_{c+x}^{j+1}(nQ^{j+1})\bigr),
$$
$$
\pi_j(\varphi)=\varphi \vert_{  Y^{j+1}},\quad
n\geq 1,\ j\geq 0,
$$
is an~epimorphism for
$x$
in general position.
}

We let
$Y^0$, ${\cal L}_c^0$,
and
$Q^0$
denote
$X^g$, ${\cal L}_c$, and~$Y$.

{\sc Proof.}
Let
$F$
be a~bundle of rank~$r$ over~$X^g$
invariant under translations.
Denote by
$F_c$
the bundle
$F\otimes{\cal P}_c$,
where
${\cal P}_c$
is the restriction of the Poincar\'e bundle to~
$X^g\times\{c\}$.
In~[1] (see Example~5.8 and Proposition~5.10)
it was proven that
$$
{\rm H}^i(X^g, F_c(nY))=0,\quad i\geq 1,\ n\geq 1,
                                             \eqno{(2)}
$$
$$
{\rm H}^i(X^g, F_c(nY))=0,
\quad i\ne g,\ n\leq -1,
                                                   \eqno{(3)}
$$
and the equality
$$
{\rm H}^i(X^g, F_c)=0\eqno{(4)}
$$
is valid for a~point
$c$
in general position for
$i\geq 0$.
Observe that the bundle
$$
{\cal L}_c\otimes[sY]\otimes[-Y_{a_1}]\otimes
\dots\otimes[-Y_{a_s}]
$$
is invariant under translations, where
$1\leq s\leq k$,
since so are
${\cal L}_c$
and
$[Y]\otimes[-Y_{a_j}]$.
Hence,
$$
{\rm H}^i(X^g,
{\cal L}_c\otimes[nY]\otimes[-Y_{a_1}]\otimes
\dots\otimes[-Y_{a_s}])=0,\eqno{(5)}
$$
where
$1\leq i<g$ and $n\in{\Bbb Z}$.
We have the exact sequence of bundles
$$
0\rightarrow{\cal L}_c^j
\otimes [nQ^j]\otimes [-Y^{j+1}]
\rightarrow{\cal L}_c^j\otimes [nQ^j]
\rightarrow{\cal L}_c^{j+1}\otimes [nQ^{j+1}]\rightarrow 0.
\eqno{(6)}
$$
It follows from
the long exact cohomology sequence corresponding to this sequence
that, for proving surjectivity of~
$\pi_j$,
it suffices to establish the equality
$$
{\rm H}^i\bigl(Y^j,
{\cal L}_c^j\otimes[nQ^j]\otimes[-Y^{j+1}]\bigr)=0.
                                                 \eqno{(7)}
$$
From (5) we immediately obtain surjectivity of~
$\pi_0$.
To prove (7), consider the following exact sequence:
$$
0\rightarrow{\cal L}_c^j \otimes [nQ^j]\otimes
[-(Y^j\cap Y_{a_{j+1}})] \otimes\dots\otimes [-(Y^j\cap
Y_{a_{j+s}})]
$$
$$
 \rightarrow{\cal L}_c^j \otimes
[nQ^j]\otimes [-(Y^j\cap Y_{a_{j+2}})] \otimes\dots\otimes
[-(Y^j\cap Y_{a_{j+s}})]
$$
$$
\rightarrow{\cal L}_c^{j+1}\otimes [nQ^{j+1}]\otimes
[-(Y^{j+1}\cap Y_{a_{j+2}})]
\otimes\dots\otimes
[-(Y^{j+1}\cap Y_{a_{j+s}})]\rightarrow 0,
\eqno{(8)}
$$
where
$j+s\leq k$.
From the long exact cohomology sequences corresponding to~(6) and
(8),
using~(2)--(4) and inducting on~$j$, we obtain
$$
{\rm H}^i\bigl(Y^j,{\cal L}_c^j\otimes [nQ^j]\otimes
[-(Y^j\cap Y_{a_{j+1}})]
\otimes\dots\otimes
[-(Y^j\cap Y_{a_{j+s}})]\bigr)=0,
                                          \eqno{(9)}
$$
where
$1\leq i<g-j$ and $j+s\leq k$.
Consequently,
$\pi_j$
is surjective.
The lemma is proven.

Observe also that if
$n>g$
then ~(9) is valid for
$i\geq 1$.

From Lemma~2 we derive

{\bf Corollary 2.} {\sl
The restriction map
$$
\pi_j: M_c^j
\rightarrow
M_c^{j+1},\quad
\pi_j(\varphi)=\varphi|_{  Y^{j+1}},
$$
is an~epimorphism for
$c$
in  general position.
}

We also need

{\bf Lemma 3.} {\sl
The linear span of the set
$$
\biggl\{\bigcup_{b, \varphi}\frac{\vartheta_{j+1}(z-b)}
{\vartheta(z)}\,\varphi,\
\varphi\in {\rm H}^0(X^g,{\cal L}_{c+sb}
((n-1)Y),\ b\in{\Bbb C}^g\biggr\},
$$
where
$n>g$
and the union is taken over all
$b$
and
$\varphi$,
coincides with
${\rm H}^0(X^g,{\cal L}_{c}(nY))$.
}

{\sc Proof.}
Consider the sequence of mappings
$$
0\longrightarrow
{\rm H}^0(X^g,{\cal L}_{c}(nY))
\stackrel{\pi_0}{\longrightarrow}
{\rm H}^0
\bigl(Y^1,{\cal L}^1_{c}(nQ^1)\bigr)
\stackrel{\pi_1}{\longrightarrow}
$$
$$
\dots \stackrel{\pi_{g-2}}{\longrightarrow}
{\rm H}^0\bigl(Y^{g-1},{\cal L}_{c}^{g-1}(nQ^{g-1})\bigr)
\longrightarrow 0.
$$
Here
$Y^{g-1}$
is the Riemann surface
$Y^{g-2}\cap Y_{a_{g-1}}$,
where
$a_{g-1}$
is some element of~$X^g$.
The mappings
$\pi_0,\dots,\pi_{g-3}$
are surjective by Lemma~2. Since (9) holds
for
$n>g$; therefore,
$\pi_{g-2}$
is also surjective.
Consequently, to prove the lemma, it  suffices
to demonstrate that the linear span of the restrictions of the
vector-functions listed in the lemma to
$Y^{g-1}$
coincides with
${\rm H}^0\bigl(Y^{g-1},{\cal L}_{c}^{g-1}(nQ^{g-1})\bigr)$.
Take
$b_1$, $b_2\in{\Bbb C}^g$
so that the divisors
$B_1=Y_{b_1}\cap Y^{g-1}$
and
$B_1=Y_{b_1}\cap Y^{g-1}$
be disjoint. Note that from (6), (8), and (9) we obtain the
equalities
$$
{\rm H}^1\bigl({\cal L}_c^{g-1}\otimes[nQ^{g-1}]\bigr)=0,
$$
$$
{\rm H}^1\bigl(Y^{g-1},
{\cal L}_c^{g-1}\otimes[nQ^{g-1}]\otimes[-B_i]\bigr)=0,
$$
$$
{\rm H}^1\bigl(Y^{g-1},
{\cal L}_c^{g-1}\otimes[nQ^{g-1}]\otimes[-B_1]\otimes[-B_2]\bigr)=0.
$$
Then by the Riemann--Roch theorem
$$
h^0\bigl({\cal L}_c^{g-1}\otimes[nQ^{g-1}]\bigr)=
\deg\bigl({\cal L}_c^{g-1}\otimes[nQ^{g-1}]\bigr)-(g(Y^{g-1})-1)r,
$$
$$
h^0\bigl({\cal L}_c^{g-1}\otimes[nQ^{g-1}]\otimes[-B_i]\bigr)=
\deg\bigl({\cal L}_c^{g-1}\otimes[nQ^{g-1}]\otimes[-B_i]\bigr)
-(g(Y^{g-1})-1)r,
$$
$$
h^0\bigl({\cal L}_c^{g-1}
\otimes[nQ^{g-1}]\otimes[-B_1]\otimes[-B_2]\bigr)
$$
$$
=\deg\bigl({\cal L}_c^{g-1}\otimes[nQ^{g-1}]\otimes[-B_1]
\otimes[-B_2]\bigr)
-(g(Y^{g-1})-1)r,
$$
where
$h^0$
is the dimension of
${\rm H}^0$ and $g(Y^{g-1})$
is the genus of~
$Y^{g-1}$.
Since the divisors~
$B_1$
and
$B_2$
are disjoint, we have
$$
\dim\bigl(
{\rm H}^0
(Y^{g-1},{\cal L}_c^{g-1}(nQ^{g-1})\otimes[-B_1]\bigr)
\cap
{\rm H}^0
\bigl(Y^{g-1},{\cal L}_c^{g-1}(nQ^{g-1})\otimes[-B_2]\bigr)\bigr)
$$
$$
=h^0\bigl({\cal L}_c^{g-1}(nQ^{g-1})\otimes[-B_1]
\otimes[-B_2]\bigr).
$$
Recall that we identify
${\rm H}^0
\bigl(Y^{g-1},{\cal L}_c^{g-1}(nQ^{g-1})\otimes[-B_j]\bigr)$
with the space of global sections of
${\cal L}_c^{g-1}(nQ^{g-1})$
having zeros at the points of the divisor~
$B_j$.
Hence, we obtain the equality
$$
h^0\bigl({\cal L}_c^{g-1}\otimes[nQ^{g-1}]\bigr)=
h^0\bigl({\cal L}_c^{g-1}\otimes[nQ^{g-1}]\otimes[-B_1]\bigr)
$$
$$
+h^0\bigl({\cal L}_c^{g-1}
\otimes[nQ^{g-1}]\otimes[-B_2]\bigr)-
h^0\bigl({\cal L}_c^{g-1}
\otimes[nQ^{g-1}]\otimes[-B_1]\otimes[-B_2]\bigr),
$$
which means that the linear span of the restriction of~
$W(b_1)\cup W(b_2)$
to
$Y^{g-1}$
coincides with
${\rm H}^0\bigl({\cal L}_c^{g-1}\otimes[nQ^{g-1}]\bigr)$,
where
$$
W(b)=
\biggl\{\frac{\vartheta_{j+1}(z-b)}
{\vartheta(z)}\,\varphi,\ \varphi\in {\rm H}^0
(X^g,{\cal L}_{c+sb}((n-1)Y)\biggr\}.
$$
This completes the proof of the lemma.

Note that we have proven even more. In the condition of Lemma~3
we need not take the union over all~$b$;
it suffices to take the set
$$
W(a_1)\cup\dots\cup W(a_{g-2})\cup W(b_1)\cup W(b_2).
$$

Denote by
$S^g_n$
the dimension of the space of differential
operators in $g$ variables with constant coefficients
whose degree does not exceed~$n-1$. It is easy to verify that
$$
S^g_n=C^{n-1}_{n+g-1}=
\frac{n(n+1)\dots (n+g-1)}{g!}.
$$
Introduce one more notation
$$
{\cal F}_j(n)=\dim {\rm H}^0
\bigl(Y^j,{\cal L}_c^j (nQ^j)\bigr),
\quad  0\leq j<g-1.
$$

{\sc Proof of Theorem~2.}
Take a~homogeneous basis
$\Phi_c$
for the ${\cal D}_g$-module~$M_c$
so that its restriction to the variety
$Y^j$
generates the ${\cal D}_j$-module~$M_c^j$;
i.e.,
$$
M_c^j=\{d_1\phi_{1,c}|_{  Y^j}
+\dots +d_{N}\phi_{N,c}|_{  Y^j},\ d_i
\in{\cal D}_g \}.
$$
This is possible by Corollary~2. By homogeneity of the basis
we mean the following. First, all elements of the basis
$\Phi_c$
are contained in~
$M_c(g)$
(this requirement is fulfilled by Nakayashiki's theorem).
And, second, if
$\phi_{1,c},\dots, \phi_{1,K}$
are the elements of the basis that belong to
$M_c(n)$, $n\leq g$,
then they generate~$M_c(n)$.
In other words,
$$
\{d_1\phi_{1,c}+\dots +d_{K}\phi_{K,c},\
d_j\in{\cal D}_g\}\cap M_c(n)=M_c(n).
$$
Denote by
$a^g_n$
the number of elements of the basis~$\Phi_c$
belonging to~$M_c(n)$
but not to~$M_c(n-1)$.
Since the basis is homogeneous and the ${\cal D}_g$-module~
$M_c$
is free, we have
$$
a_1^gS_n^g+\dots+a_g^gS_{n-g+1}^g={\cal F}_0(n), \quad n>g.
$$
Denote by
$
{\cal D}_{g-j}\Phi_c^j\subset M_c^j
$
the ${\cal D}_{g-j}$-module
$$
\{\varphi|_{  Y^j},\ \varphi=d_1\phi_{1,c}+\dots
+d_{N}\phi_{N,c},\ d_s\in{\cal D}_{g-j}\}.
$$
Inducting on~$k$, prove that
$
{\cal D}_{g-k}\Phi_c^k
$
is a~free ${\cal D}_{g-k}$-module of rank~$N$.
Then, computing the dimensions of the spaces~
$M_c^j(n)$
and
${\cal D}_{g-j}\Phi_c^j\cap M_c^j(n)$
(for a~fixed~$x$),
we establish that these ${\cal D}_{g-k}$-modules coincide.

The initial induction step is Nakayashiki's theorem.
Suppose that the assertion is proven for
$k=j$.
Since the ${\cal D}_{g-j}$-module~
$M_c^j$
is free, we derive the equality
$$
a_1^gS_n^{g-j}+\dots+a_g^gS_{n-g+1}^{g-j}={\cal F}_j(n),\quad
n>g.
                                         \eqno{(10)}
$$
Suppose that the ${\cal D}_{g-j-1}$-module
${\cal D}_{g-j-1}\Phi_c^{j+1}$
is not free for
$k=j+1<g$.
Then there exist operators
$\tilde{d}_i\in{\cal D}_{g-j-1}$
such that
$$
\phi=\tilde{d}_1\phi_{1,c}+\dots
+\tilde{d}_{N}\phi_{N,c},\quad
\phi|_{  Y^{j+1}}=0.
$$
This is equivalent to the fact that
$ M_c^j(n) $
(we may assume that
$n>g$)
contains an~element of the form
$\frac{\vartheta_j(z-a_{j+1})}{\vartheta(z)}\varphi$,
$\varphi\in M_{c-a_{j+1}}^j(n-1)$,
for which
$$
\tilde{d}_1\phi_{1,c}+\dots
+\tilde{d}_{N}\phi_{N,c}=
\frac{\vartheta_{j+1}(z-a_{j+1})}{\vartheta(z)}\varphi,
\quad z\in Y^j.
\eqno{(11)}
$$
Consider the following subspace in
${\rm H}^0\bigl(Y^j, {\cal L}_{c+x}^j(nQ^j)\bigr)$:
$$
V_{c+x}^j(n)=
\biggl\{
\frac{\psi}{e}\bigg\vert_{  Y^j},\
\psi=d_1\phi_{1,c}+\dots+d_{N,c}\phi_{N},\ d_i\in{\cal D}_{g-j-1}
\biggr\}
\cap {\rm H}^0\bigl(Y^j, {\cal L}_{c+x}^j(nQ^j)\bigr),
$$
where
$$
e=\exp\Biggl(-\sum\limits_{j=1}^g
\frac{x_j}{s}\,\partial_{z_j}\log\vartheta(z)\Biggr).
$$
Find the dimension of
$
V_{c+x}^j(n).
$
From (10) we obtain the equality
$$
a_1^g\bigl(S_n^{g-j}-S_{n-1}^{g-j}\bigr)+\dots+
a_g^g\bigl(S_{n-g+1}^{g-j}-S_{n-g}^{g-j}\bigr)=
a_1^gS_n^{g-j-1}+\dots+
a_g^gS_{n-g+1}^{g-j-1}
$$
$$
={\cal F}_j(n)-{\cal F}_j(n-1)={\cal F}_{j+1}(n);
$$
consequently,
$$
\dim V_{c+x}^j(n)={\cal F}_j(n)-{\cal F}_j(n-1).\eqno{(12)}
$$
Introduce one more subspace in
${\rm H}^0\bigl(Y^j,{\cal L}_{c+x}^j(nQ^j)\bigr)$
which depends on the element
$a_{j+1}$:
$$
W_{c+x}^j(n)=
\biggl\{
\frac{\vartheta_{j+1}(z-a_{j+1})}{\vartheta(z)}\,\varphi,\
\varphi\in {\rm H}^0
\bigl(Y^j, {\cal L}_{c+x+sa_{j+1}}^j((n-1)Q^j\bigr),\
z\in Y^j
\biggr\}.
$$
It is clear that
$$
\dim W_{c+x}^j(n)={\cal F}_j(n-1),
$$
since $\dim {\rm H}^0\bigl(Y^j,
{\cal L}_{c+x+sa_{j+1}}^j((n-1)Q^j)\bigr)={\cal F}_j(n-1)$.

It follows from Lemmas~2 and~3 that there is an~element
$a_{j+1}$
for which an~equality like ~(11) is impossible. Since
$$
\dim V_{c+x}^j(n)
+\dim W_{c+x}^j(n)
=\dim {\rm H}^0\bigl(Y^j,
{\cal L}_{c+x}^j(nQ^j)\bigr),
$$
an equality like~(11) is impossible for elements of some
small neighborhood of~
$a_{j+1}$.
By analytic dependence of the space
$W_{c+x}^j(n)$
on~
$a_{j+1}$,
an~equality like~(11) does not hold
for an~open everywhere dense set of such
$a_{j+1}$.
Consequently, since the set
$a_1,\dots,a_k$
is in general position by assumption,
${\cal D}_{g-j-1}\Phi_c^{j+1}$
is a~free ${\cal D}_{g-j-1}$-module of rank~$N$.

Prove that the ${\cal D}_{g-j-1}$-modules
${\cal D}_{g-j-1}\Phi_c^{j+1} $
and $M_c^{j+1}$
coincide.

Since
$\frac{\vartheta_j(z)}{\vartheta(z)}$
is a~meromorphic function and sequence~(6) is exact,
we obtain the equality
$$
\dim {\rm H}^0\bigl(Y^{j+1}, {\cal L}_{c+x}^{j+1}(nQ^{j+1})\bigr)
$$
$$
=\dim {\rm H}^0\bigl(Y^j, {\cal L}_{c+x}^j(nQ^j)\bigr)
-\dim {\rm H}^0\bigl(Y^j,
{\cal L}_{c+x}^j((n)Q^j)\otimes[-Y^{j+1}]\bigr)
$$
$$
=\dim {\rm H}^0\bigl(Y^j, {\cal L}_{c+x}^j(nQ^j)\bigr)
-\dim {\rm H}^0\bigl(Y^j,
{\cal L}_{c+x}^j((n-1)Q^j)\bigr)={\cal F}_j(n)-{\cal F}_j(n-1).
$$
Observing the inclusion
$ {\cal D}_{g-j-1}\Phi_c^{j+1}\subset M_c^{j+1}$,
from~(12) we find
$$
\dim {\rm H}^0\bigl(Y^{j+1},
{\cal L}_{c+x}^{j+1}(nQ^{j+1})\bigr)=
\dim V_{c+x}^j(n)={\cal F}_{j+1}(n).
$$
So the ${\cal D}_{g-j-1}$-modules
${\cal D}_{g-j-1}\Phi_c^{j+1} $
and
$M_c^{j+1}$
coincide.
Theorem~3 is proven.

We now demonstrate how to derive Theorems~1 and~2 from Theorem~3.
Denote by
$$
\Phi_c=(\phi_{1,c}(z,x),\dots,\phi_{N,c}(z,x))^{\top}
$$
a~basis for the ${\cal D}_{g-k}$-module~
$M_c^k$. Then, by Theorem~3, for
$\lambda\in {\cal A}_k$,
there is a~unique operator
$L_k(\lambda)\in  {\rm Mat}(N,g-k)$
such that
$$
L_k(\lambda)\Phi_c=\lambda\Phi_c.
$$
The operators
$L_k(\lambda)$
with different
$\lambda$'s
commute pairwise.
Theorem~1 is proven.

Denote by
$T_j\in {\rm Mat}(N,g-k)$
the operator of order~
$g$
defined by the equality
$$
T_j\Phi_c=\partial_{t_j}\Phi_c;
$$
here we identify time
$t_j$, $1 \leq j\leq k$,
with the variable
$x_{g-k}$.
The equalities
$$
[L_k(\lambda),T_j-\partial_{t_j}]\Phi=0,\quad
[T_m-\partial_{t_m},T_n-\partial_{t_n}]\Phi=0
$$
hold. Then Theorem~3 yields

{\bf Corollary 3.} {\sl
The following evolution equations are valid:
$$
\frac{\partial L_k(\lambda)}{\partial t_j}=
[L_k(\lambda),T_j],
\quad  \lambda\in {\cal A}_k,
$$
$$
\frac{\partial T_m}{\partial t_n}-
\frac{\partial T_n}{\partial t_m}
=
[T_n,T_m].
$$
}

We turn to proving Theorem~2.
We divide each vector-function~
$\phi_{j,c}(z,x)$
by
$$
\exp
\Biggl(
-\sum\limits_{j=g-k+1}^g
\frac{x_j}{s}\,\partial_{z_j}\log\vartheta(z)
\Biggr),
$$
and then replace
$x=(x_1,\dots,x_g)$
with
$$
(x,t)=\Bigl(x_1,\dots,x_{g-k},
\sum\limits_mt_{1,m},\dots,\sum\limits_mt_{k,m}\Bigr),
$$
where
$m=(m_1,\dots,m_g)\in{\Bbb Z}^g$,
$m_1+\dots+m_g\geq 2$,
$m_i\geq 0$,
and, finally, multiply by
$$
\exp
\Biggl(
-\sum\limits_{j=1}^{k}\sum\limits_m
\frac{t_{j,m}}{s}\bigl(\partial_{z_{g-k+j}}\log\vartheta(z)+
\partial_z^m\log\vartheta(z)\bigr)
\Biggr),
$$
where
$
\partial_z^m\log\vartheta(z)=\partial_{z_1}^{m_1}\dots
\partial_{z_g}^{m_g}\log\vartheta(z).
$
We obtain the vector-function
$\psi_{j,c}(z,x,t)$
which is representable as the sum of vector-functions:
$$
  g(x,t)\frac{\theta^{r,sn}\bigl(z+\frac{(x,t)+c}{sn}\bigr)}
{\vartheta^n(z)}
\exp\Biggl(-\sum\limits_{j=1}^{g-k}\frac{x_j}{s}
\partial_{z_j}\log\vartheta(z)
$$
$$
-\sum\limits_{j=1}^{k}\sum\limits_m
\frac{t_{j,m}}{s}\bigl(\partial_{z_{g-k+j}}\log\vartheta(z)+
\partial_z^m\log\vartheta(z)\bigr)
\Biggr).
$$
Then for
$$
\Psi=(\psi_{1,c}(z,x,t),\dots,\psi_{N,c}(z,x,t))^{\top}
$$
we have the equality
$$
L_{j,m}\Psi=\partial_{t_{j,m}}\Psi.
$$
By Theorem~3,
$$
[\partial_{t_{j,m}}-L_{j,m},
\partial_{t_{i,n}}-L_{i,n}]=0.
$$
Theorem~2 is proven.

{\bf References.}

[1] Nakayashiki~A. Commuting partial differential operators
and vector  bundles over abelian  varieties.
      Amer.~J. Math., Vol. 116 (1994), 65--100.

[2] Nakayashiki~A.
      Structure of Baker--Akhiezer modules of
principally polarized abelian varieties, commuting partial
differential operators and associated integrable systems.
       Duke Math.~J., Vol. 62 (1991), N. 2, 315--358.

[3] Mironov A.~E. Commutative rings of~differential operators
connected with~two-dimensional abelian  varieties.
Siberian Math. J., Vol. 41 (2000), N. 6, 1148--1161.

[4]
Mironov A.~E.
Real commuting differential operators connected
with~two-dimensional abelian varieties
Siberian Math. J., Vol. 43 (2002), N. 1, 97--113

[5]
Rothstein~M. Sheaves with connection on abelian varieties.
Duke Math.~ J.,  Vol.    84 (1996), N. 3, 565--598.

[6]
 Krichever~I.~M.
The methods of algebraical geometry
      in the theory of nonlinear equations. Uspekhi Mat. Nauk.
     Vol. 32 (1997), N. 6, 183--208.

[7]  Mukai~S. Duality between $D(X)$ and $D(\widehat{X})$ with
its application to Picard sheaves. Nagoya Math.~J.,
 Vol.  81 (1981), 153--175.

\vskip3mm
{\sl
Sobolev Institute of Mathematics,
Novosibirsk}

  mironov@math.nsc.ru

\enddocument